\documentclass[aps,nature,preprint,superscriptaddress,longbibliography]{revtex4-1}

\usepackage[utf8]{inputenc}
\usepackage{amsmath}
\usepackage{color}
\usepackage{verbatim}
\usepackage{amsmath}
\usepackage{amssymb}
\usepackage{graphicx}
\usepackage{dirtytalk}
\usepackage{verbatim}
\usepackage{grffile}
\usepackage{dcolumn}
\usepackage{bm}
\usepackage[dvipsnames]{xcolor}
\usepackage{siunitx}
\usepackage{braket}
\usepackage[normalem]{ulem}
\usepackage{lineno} 

\usepackage[colorlinks=true,allcolors=blue]{hyperref}

\begin{document}

\title{Collective flow of fermionic impurities immersed in a Bose-Einstein Condensate}
\author{Zoe Z.~Yan}
\affiliation{
MIT-Harvard Center for Ultracold Atoms, Research Laboratory of Electronics, and Department of Physics,
 Massachusetts Institute of Technology, Cambridge, Massachusetts 02139, USA }
\affiliation{Physics Department and James Franck Institute, University of Chicago, Chicago, Illinois, USA}
\author{Yiqi Ni}
\affiliation{
MIT-Harvard Center for Ultracold Atoms, Research Laboratory of Electronics, and Department of Physics,
 Massachusetts Institute of Technology, Cambridge, Massachusetts 02139, USA }
\author{Alexander Chuang}
\affiliation{
MIT-Harvard Center for Ultracold Atoms, Research Laboratory of Electronics, and Department of Physics,
 Massachusetts Institute of Technology, Cambridge, Massachusetts 02139, USA }
\author{Pavel E.~Dolgirev}
\affiliation{
 Department of Physics, Harvard University, Cambridge, Massachusetts 02138, USA}
\affiliation{
Institute for Theoretical Physics, ETH Zurich, 8093 Zurich, Switzerland}
\author{Kushal Seetharam}
\affiliation{
 Department of Physics, Harvard University, Cambridge, Massachusetts 02138, USA}
 \affiliation{
Department of Electrical Engineering, Massachusetts Institute of Technology, Cambridge, Massachusetts 02139, USA }
\author{Eugene Demler}
\affiliation{
Institute for Theoretical Physics, ETH Zurich, 8093 Zurich, Switzerland}
\author{Carsten Robens}
\affiliation{
MIT-Harvard Center for Ultracold Atoms, Research Laboratory of Electronics, and Department of Physics,
 Massachusetts Institute of Technology, Cambridge, Massachusetts 02139, USA }
\author{Martin Zwierlein}
\altaffiliation{Email: zwierlei@mit.edu}
\affiliation{
MIT-Harvard Center for Ultracold Atoms, Research Laboratory of Electronics, and Department of Physics,
 Massachusetts Institute of Technology, Cambridge, Massachusetts 02139, USA }

\date{\today}

\maketitle


\textbf{
Interacting mixtures of bosons and fermions are ubiquitous in nature. They form the backbone of the standard model of physics, provide a framework for understanding quantum materials, and are of technological importance in helium dilution refrigerators. However, the description of their coupled thermodynamics and collective behavior is challenging. Bose-Fermi mixtures of ultracold atoms provide a platform to investigate their properties in a highly controllable environment, where the species concentration and interaction strength can be tuned at will. Here, we characterize the collective oscillations of spin-polarized fermionic impurities immersed in a Bose-Einstein condensate as a function of the interaction strength and temperature. For strong interactions, the Fermi gas perfectly mimics the superfluid hydrodynamic modes of the condensate, from low energy quadrupole modes to high order Faraday excitations. With an increasing number of bosonic thermal excitations, the dynamics of the impurities cross over from the collisionless to the hydrodynamic regime, reminiscent of the emergence of hydrodynamics in two-dimensional electron fluids.}

The paradigmatic example of fermions coupled to a bosonic bath is the motion of itinerant electrons through an ionic crystal. The coupling to the ionic lattice vibrations endows the electrons with a shifted energy and mass, as they become dressed into polarons~\cite{Landau1933, Pekar1946}, the first instance of the quasiparticle concept. We also encounter Bose-Fermi mixtures as dilute solutions of fermionic $^3$He in bosonic superfluid $^4$He~\cite{Ebner1971}, in quark-meson models in high-energy physics~\cite{schaefer_phase_2005} and in 2D electronic materials, where interactions between excitons and electrons can be controlled~\cite{sidler2017fermi, Shimazaki2020, Schwartz2021}.
Ultracold atomic gases provide arguably the purest realizations of Bose-Fermi mixtures, featuring precisely understood, tunable short-range interactions and a high degree of experimental control~\cite{Hadzibabic2002,Stan2004,Inouye2004,Silber2005,Ospelkaus2006,Zaccanti2006,Shin2008,Wu2011,Park2012,Vaidya2015,Trautmann2018,Lous2018,DeSalvo2019,patel2022observation}, offering a direct comparison to theoretical models
~\cite{Viverit2000,Buchler2003,Bertaina2013,Kinnunen2015,Ludwig2011}.
In recent years, atomic Bose-Fermi mixtures enabled the study of dual superfluids~\cite{Ferrier-Barbut2014,Delehaye2015,Yao2016,Roy2017,Wu2018}, the onset of phase separation and mean-field collapse~\cite{Ospelkaus2006a, Ospelkaus2006,Zaccanti2006,Lous2018,Huang2019}, and the observation of strong-coupling Bose polarons~\cite{Hu2016,Yan2020}.

The general dynamics of fermions interacting with a partially condensed Bose gas at finite temperature are challenging to describe, as interactions between fermions and bosons come in two flavors. On one hand, fermions can incoherently scatter with thermal bosons, leading to momentum-changing collisions.
On the other hand, fermions also experience momentum-preserving interactions, in particular with the BEC, in the form of an effective potential~\cite{Landau1941,Viverit2000,Chevy2015,Seetharam2021,Seetharam2021long}.
The interplay between the two types of interactions dictates the dynamics of the whole system~\cite{Miyakawa2000, Yip2001,Sogo2002,Liu2003,Capuzzi2004,Imambekov2006, Banerjee2009,VanSchaeybroeck2009,Maruyama2014, Asano2019b,Ono2019}.
Most challenging is the regime for strong interactions where the non-superfluid system crosses over from collisionless to the collisionally hydrodynamic regime. Such a regime is observed in electron-phonon mixtures in the context of high-temperature superconductivity~\cite{Zhang2017,Gooth2018}.
With such strong interactions and with bosons partially condensed, the question arises whether the system indeed remains superfluid, with fermions flowing without dissipation through the condensate. Also, it is unclear how thermal bosonic excitations alter the transport properties of fermions.
Here, we thoroughly address these questions, observing collective flow of fermionic impurities mimicking closely the superfluid hydrodynamic modes of the condensate. Despite strong interspecies interactions the flow can be modelled as collisionless and driven by the condensate's mean-field. With increasing temperature we observe a crossover to collision-dominated hydrodynamic flow.

Collective excitations are exquisitely sensitive probes of interparticle scattering and interactions.
They have been used to demonstrate the superfluid hydrodynamic flow of BECs~\cite{Jin1996,Mewes1996, Stamper-Kurn1998} and collisional hydrodynamics in interacting Fermi gases~\cite{Gensemer2001,OHara2002,Regal2003,Bourdel2003, Trenkwalder2011,Tey2013,Ravensbergen2020}.
Collective oscillations have also been measured in coupled Bose-Fermi superfluids~\cite{Ferrier-Barbut2014,Delehaye2015,Roy2017,Wu2018} and in Bose-Fermi mixtures~\cite{Ferlaino2003,Fukuhara2009,DeSalvo2019,Huang2019,patel2022observation}, revealing phenomena such as collisional hydrodynamics in thermal mixtures~\cite{Ferlaino2003}, collisionless uncoupled dipole oscillations in degenerate mixtures~\cite{Ferlaino2003}, and sound propagation~\cite{patel2022observation}.
Here we study a novel regime -- the limit of a dilute gas of spin-polarized fermions immersed in a BEC -- as relevant for the physics of Bose polarons~\cite{Hu2016,Yan2020} and unconventional superconductors with low carrier densities~\cite{kagan2019fermi}.

\begin{figure}[t]
	\includegraphics[width=\textwidth]{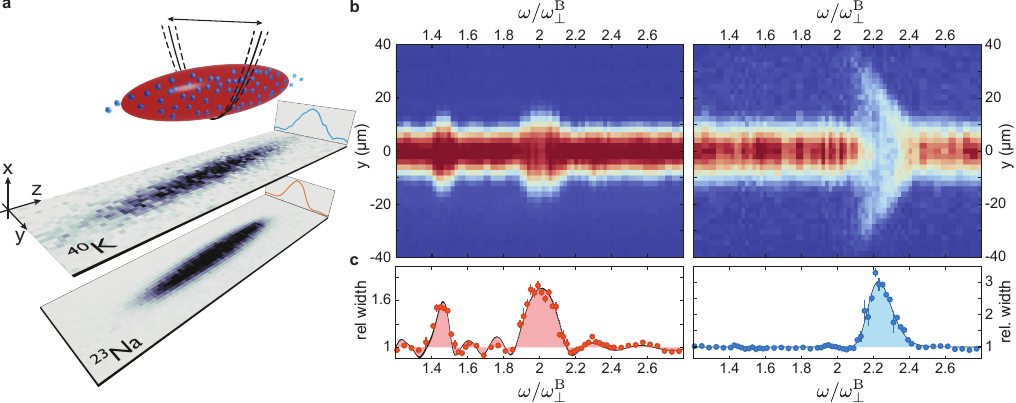}
	\caption{\label{fig:schematic} \textbf{Collective oscillations in a Bose-Fermi mixture.  a.} 
	Illustration of a dilute gas of fermions (blue) immersed in a Bose-Einstein condensate (red), both trapped in an optical potential. \textit{In-situ} absorption images of the fermionic $^{40}$K and bosonic $^{23}$Na are shown beneath. Radial collective oscillations are induced by periodically modulating the depth of the optical potential. 
	\textbf{b.} The doubly integrated line densities along the transverse (y) direction of the boson (left) and fermion (right) clouds as a function of modulation frequency $\omega$, at zero interspecies interaction ($a_\mathrm{BF}\,{=}\,0$). 
	\textbf{c}.  Spectra of the uncoupled Bose-Fermi system, showing the boson (left, red circles) and fermion (right, blue circles) relative widths, reveal the superfluid hydrodynamic response of the BEC and the response of the non-interacting, near-degenerate gas of spin-polarized fermions.  The solid black lines show predictions from the hydrodynamic scaling ansatz in Eq.~\ref{eqn:scalingB} and a phenomenological Gaussian fit, respectively.  All error bars show the standard error of the mean.
	}
\end{figure}

We probe collective excitations of $^{40}$K fermions and a $^{23}$Na BEC as a function of drive frequency and across a range of Bose-Fermi interactions, revealing the energy and spectral width of low-lying excitations. 
The experiment starts with an ultracold near-degenerate gas of fermionic ``impurity'' atoms immersed in the BEC,  both held in a \SI{1064}{\nano\meter} crossed optical dipole trap with near-cylindrical symmetry (see Fig.~\ref{fig:schematic}a).
For our coldest samples we evaporatively cool both species to $T{\approx}$\,30\,nK, corresponding to $T/T_\mathrm{c}\,{\lesssim}\,0.2$ and $T/T_\mathrm{F}\,{\approx}\,0.6$, where $T_\mathrm{c(F)}$ is the condensate's critical temperature (the Fermi temperature).
Both species are in their respective hyperfine ground states.
We control the interspecies interactions by ramping the magnetic field near Feshbach resonances, allowing us to continuously tune the $s$-wave scattering length, $a_\mathrm{BF}$.
The typical peak boson density is $n_\mathrm{B}\,{=}\,7\times 10^{13}$\,cm$^{-3}$, and the typical impurity concentration varies between $n_\mathrm{F}/n_\mathrm{B}\,{\approx}\,0.003-0.02$ (see Methods).

We characterize the low-energy radial excitations of the mixture at varying interspecies interaction strengths. 
Modulating the radial trapping potential depth at frequency $\omega$ for 10 cycles, we measure the \textit{in situ} width of the clouds (see Fig.~\ref{fig:schematic}a.) 
The number of cycles $N$ allows for spectral resolution ${\sim}\,\omega/N$ while the probe time is kept short compared to the mixture's lifetime, limited by three-body loss.
When the modulation excites a resonant mode of either species, the cloud's width expands radially.
Fig.~\ref{fig:schematic}b-c depicts the bosonic and fermionic spectrograms for a decoupled mixture at $a_{\rm BF}{=}\,0$.
Two resonances are observed in the BEC at $\sqrt{2}\omega_\perp^{\text{B}}$ and $2\omega_\perp^{\text{B}}$, 
while only one fermionic resonance is excited at $2\,\omega_y^{\text{F}}$.
Here, $\omega^\mathrm{B}_\perp\,{=}\,(\omega^\mathrm{B}_x\omega^\mathrm{B}_y)^{1/2}$ is the bosons' geometric mean radial trapping frequency, and $\omega_y^{\text{F}}$ is the fermions' trap frequency along the transverse $y$ direction, with the two frequencies related by $\omega_y^{\text{F}}=1.16\,\omega_\perp^{\text{B}}$ ~\cite{Supplement}.
The BEC obeys superfluid hydrodynamics, which couples the two collisionless radial modes, giving rise to a quadrupole (out of phase) and a breathing (in phase) resonance at $\sqrt{2}\omega_\perp^{\text{B}}$ and $2\omega_\perp^{\text{B}}$, respectively~\cite{Stringari1996,Pethick2008}.
Across all frequencies, the bosonic spectral response is well captured using a hydrodynamic scaling ansatz~\cite{Castin1996,Supplement} (see red shaded area in Fig.~\ref{fig:schematic}c):

\begin{align}
    \ddot{b}_i&+{\omega_{i}^B}(t)^2b_i-\frac{{\omega_{i}^B}(0)^2}{b_i\prod_jb_j}=0, \label{eqn:scalingB}
\end{align}
where $i\,{\in}\,(x,y,x)$ and $b_i$ is the dimensionless scaling parameter of the BEC's Thomas-Fermi radius in the $i$-th direction.
The spin-polarized fermions are a collisionless gas~\cite{DeMarco1999}, which in a purely harmonic trap has its lowest parametric resonance for motion along the y-axis at $2\omega_y^{\text{F}}$. 
The energy and spectral width of the fermions' spectral response is obtained using a fit to a phenomenological function~\cite{Supplement} (see Fig.~\ref{fig:schematic}c).
The broad fan below $2\omega_y^{\text{F}}$ visible in the fermionic response (Fig.~\ref{fig:schematic}b) arises from strong driving in an anharmonic trap~\cite{Supplement}.

\begin{figure}[t]
\includegraphics[width=0.5\columnwidth]{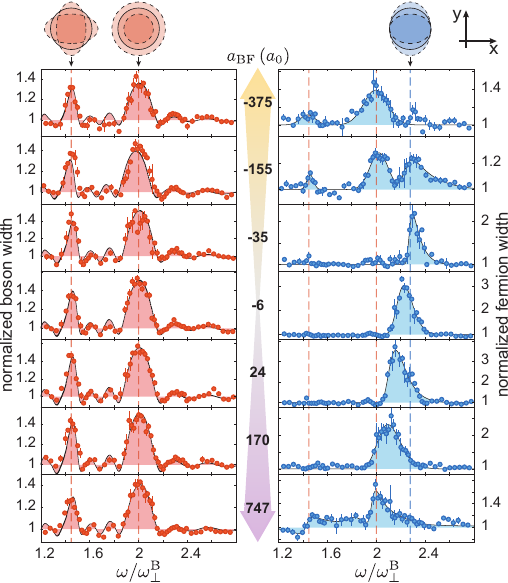}
	\caption{\label{fig:collective1} 
	\textbf{Evolution of Bose-Fermi collective modes across varying interaction strength.} Illustrations depict the oscillations of the cloud's transverse ($x$-$y$) cross-section.  The boson quadrupole and breathing modes (shown in red) lie at $\sqrt{2}\omega^\mathrm{B}_\perp$ and $2\omega^\mathrm{B}_\perp$, respectively, whereas the fermions' transverse resonance in the collisionless regime lies at $2\omega^\mathrm{F}_y$ (shown in blue).
	 The spectra depict bosonic and fermionic cloud widths (circles) for varying modulation frequencies. As $a_{\rm BF}$ increases toward either repulsive or attractive interactions, the fermion spectra evolve to mode lock to the BEC's superfluid hydrodynamic modes. An extended data set can be found in ~\cite{Supplement}.
	}	
\end{figure}

Fig.~\ref{fig:collective1} shows a selection of the bosons' and fermions' spectral response for various interspecies coupling strengths.
The resonances of the BEC are always well described by the hydrodynamic scaling ansatz from Eq.~\ref{eqn:scalingB}.
The BEC spectrogram is unaffected by interactions with the much more dilute gas of fermionic impurities.
By contrast, the response of the fermions shows a strong dependence on the interspecies coupling strength.
At weak couplings, the collisionless mode of the fermions shifts linearly with $a_\text{BF}$.
With increasing interactions strengths, the fermionic response changes drastically, revealing two additional modes in the their spectral response that coincide with the BEC's hydrodynamic superfluid modes at $2\omega_\perp^{\text{B}}$ and $\sqrt{2}\omega_\perp^{\text{B}}$. 
All three modes are spectrally well resolved and show no broadening beyond the Fourier limit, in contrast to the broadened profiles that would arise from momentum-relaxing collisions~\cite{Supplement}. This is remarkable, given that the mean-free path for collisions changes from infinity at zero interaction strength to $l_{\rm mfp} = (4\pi a_\mathrm{BF}^2n_\mathrm{B})^{-1}\,\sim \,0.6\,\mu$m at the strongest measured interactions, much shorter than the radial system size $L\sim 10\,\mu$m. 
A thermal mixture would thus cross over from collisionless to collisionally hydrodynamic behavior through an intermediate regime of strong damping. Here, instead, 
the fermions remain collisionless with the condensed bosons,
and, for the strongest interactions, even ``copy'' the BEC's superfluid collective modes, not unlike dye particles in water.

\begin{figure}[t]
		
	\includegraphics[width=0.5\columnwidth]{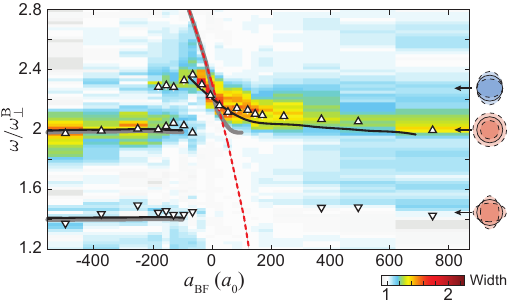}
	\caption{\label{fig:collective2} 
   \textbf{Fermionic mode frequencies versus interspecies interaction.} The color density plot displays the power spectra of cloud widths. White markers denote the peak frequencies. The arrows on the right side indicate the three modes of Fig.~\ref{fig:collective1}. The dashed red line shows the naive mean-field prediction, and the grey lines are the scaling ansatz solution of the mean-field model Eq.~\ref{eqn:scalingFm}. The black lines show the dominant modes of the collisionless Boltzmann-Vlasov solution accounting for finite system size and trap anharmonicity~\cite{Supplement}.  
	}
\end{figure}

The fermion collective modes are summarized in Fig.~\ref{fig:collective2}.
For scattering lengths beyond $|a_\text{BF}|\,{>}\,350\,a_0$ (with $a_0$ being the Bohr radius), the fermions exclusively respond at the BEC's hydrodynamic superfluid modes and show no signal of their own collisionless mode.
For weaker interactions, we observe a mean-field like shift of the fermion frequency proportional to the sign of the interaction, which for repulsive interactions merges with the BEC's breathing mode at $2\omega_\perp^{\text{B}}$ and becomes spectrally indistinguishable. 
We note that at repulsive interactions above 170\,$a_0$ we observe a dispersive - rather than absorptive - feature in the fermionic response at the BEC quadrupole mode (Fig.~\ref{fig:collective1} and extended dataset \cite{Supplement}). This Fano-type behavior can be understood from coherent coupling of the fermionic and bosonic mode \cite{theory_paper}.
At interactions stronger than measured here, $\sim$900\,$a_0$, phase separation is predicted to occur~\cite{Viverit2000,Supplement}.

To understand the dynamics across all interaction strengths, we compare various numerical models to the data. The linear dependence of the fermionic collisionless mode at weak interaction strengths qualitatively agrees with a mean-field description, considering the effective potential experienced by the fermions immersed in the Bose gas.
The BEC in the Thomas-Fermi approximation takes on the shape of the inverted optical potential. 
The fermions thus experience a joint effective potential comprising the optical trap and the mean-field potential of the BEC, where attractive (repulsive) interspecies interactions provide a steeper 
(shallower) potential that shifts the trapping frequencies according to $\tilde{\omega}^2{=}\,{\omega^F_y}^2\left(1\,{-}\,\frac{g_{\rm BF}\alpha_{\rm B}}{g_{\rm BB}\alpha_{\rm F}}\right)$~\cite{Supplement}.
Here, $g_\mathrm{BF}\,{=}\,2\pi\hbar^2a_\mathrm{BF}/m_{\rm red}$ is the Bose-Fermi coupling strength, $m_{\rm red}\,{=}\,m_\mathrm{F}m_\mathrm{B}/(m_\mathrm{F}+m_\mathrm{B})$ is the reduced mass,  $g_\mathrm{BB}\,{=}\,4\pi\hbar^2a_\mathrm{BB}/m_\mathrm{B}$ is the Bose-Bose coupling, and $\alpha_{\rm B(F)}$ is the boson (fermion) optical polarizability.
Qualitatively, this mean-field model (red dashed line in Fig.~\ref{fig:collective2}) shares the trend of the measurements for small $a_\text{BF}$, but with a different slope.
It also fails to predict the appearance of additional modes in the fermionic spectral response.
To capture the resonances that fermions inherit from the BEC's superfluid hydrodynamic modes, the mean-field potential itself must properly incorporate the bosons' response given by the scaling ansatz Eq.~\ref{eqn:scalingB}.

We therefore turn to the full dynamics of the fermions as described by the Boltzmann-Vlasov equation,
\begin{equation}
    \frac{\partial f}{\partial t}+ \frac{d \textbf{r}}{d t} \frac{\partial f}{\partial \textbf{r}}+\frac{d \textbf{p}}{d t} \frac{\partial f}{\partial \textbf{p}} = I_{\rm coll}
    \label{eq:BVE}
\end{equation}
where $f$ is the fermion distribution at momentum \textbf{p} and position \textbf{r}, and $I_{\rm coll}$ is the collision integral.
Anticipating the absence of collisions for fermions only interacting with the BEC, in the absence of thermal excitations, we set $I_{\rm coll} =0$. 
Then, to first order in $g_{\rm BF}$, we derive a scaling ansatz for the fermions' width, assuming a harmonic trap and fermionic impurities that are deeply immersed in the BEC~\cite{Castin1996,Menotti2002,Liu2003}:

\begin{align}
    \ddot{c}_i&+{{\omega^F_{\rm i,eff}}}(t)^2c_i-\left(1-\frac{g_{\rm BF}\alpha_{\rm B}}{g_{\rm BB}\alpha_{\rm F}}\right)\frac{{\omega^F_y}(0)^2}{c_i^3}=0\nonumber\\
    {\mathrm{with~} {\omega^F_{\rm i,eff}}}&(t)^2=\left(1-\frac{g_{\rm BF}\alpha_{\rm B}}{g_{\rm BB}\alpha_{\rm F}}\right){\omega^F_i}(t)^2-\frac{g_{\rm BF}}{g_{\rm BB}}\frac{\ddot{b_i}}{b_i}
    \label{eqn:scalingFm}
\end{align}
Here, $c_i$ is the dimensionless scaling parameter of the width of the gas of fermionic impurities, with $c_i(t{=}0)\,{=}\,1$.
This ansatz (shown in grey lines) captures the fermionic response to the BEC's superfluid mode on the attractive side~\cite{Supplement}, which can thus indeed be explained as the collisionless flow of fermions experiencing the coherent interactions with the BEC. The simple ansatz fails for repulsive interactions above $a_\text{BF}\,{>}\,100\,a_0$ where the mean-field potential is strong enough to repel fermions from the BEC.
A full numerical simulation of the collisionless Boltzmann-Vlasov equation~\cite{Supplement} -- including the temperature-dependent fermionic cloud size and the trap anharmonicity -- is shown as the black solid line in Fig.~\ref{fig:collective2}, which accurately captures all of the observed modes across all interaction strengths, validating our neglect of the collisional term in Eq.~\ref{eq:BVE}.

\begin{figure}
	\includegraphics[width=0.5\columnwidth]{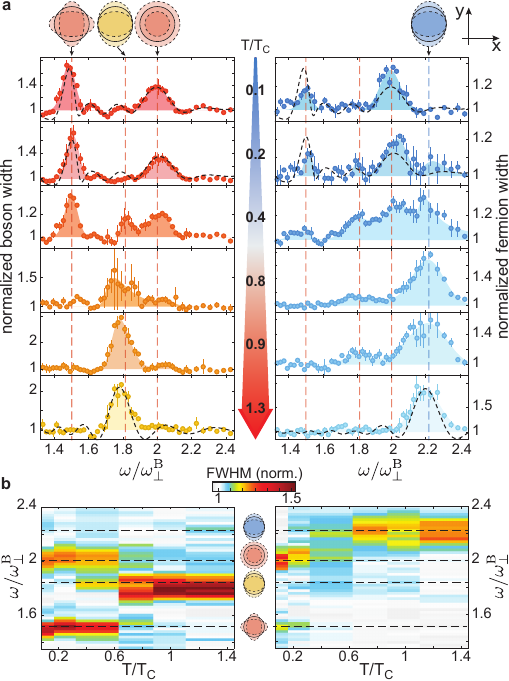}

	\caption{\label{fig:temp} \textbf{
	Temperature dependence of the bosonic and fermionic collective modes at $\mathbf{a_\mathrm{BF}\,{=}\,{-}400a_0}$. a.} 
    Spectrograms of the boson (left) and fermion (right) widths, as a function of the bosons' reduced temperature $T/T_\mathrm{c}$.  
    Top cartoons illustrate the bosonic hydrodynamic (red), collisionless (yellow), and the fermionic collisionless (blue) modes. 
	Peak boson densities for increasing $T/T_\mathrm{c}$ are $n_{\rm B}{=}(7.3,5.9,5.4,2.3,1.4,0.07){\times} 10^{13}$\,cm$^{-3}$, respectively.	The fermion spectrograms indicate hydrodynamic response at the BEC's quadrupole and breathing mode frequencies for the lowest temperatures. 
     Dashed lines in the upmost panels for $T/T_\mathrm{c} \approx 0.1$ and $T/T_\mathrm{c} \approx 0.2$ represent the theoretical prediction based on the scaling ansatz for the BEC dynamics and Boltzmann-Vlasov equation for the dynamics of fermions. Dashed lines for $T>T_\mathrm{c}$ show the result of solving coupled Bose-Fermi Boltzmann equations~\cite{Supplement}.
	\textbf{b.} Summary of the obtained peak frequencies for bosons (left) and fermions (right), with color indicating spectral response.  
	}
	
\end{figure}

Indeed, collisionless flow is expected for the impurities well below the bosons' superfluid transition temperature, as fermions slower than the condensate's speed of sound ($\sim\,5\mu$m/ms) can only dissipate energy through collisions with thermal bosons, which are essentially absent at low temperatures $T/T_\mathrm{c}\approx0.2$.
To measure the impact of collisions of fermions with thermal bosons, we now probe the mixture at increasing temperatures across the BEC phase transition. The physics is complex already for bosons alone, as the BEC becomes immersed in a cloud of thermal excitations, and the thermal cloud's collective modes couple to those of the superfluid~\cite{Pethick2008}. 
We employ the same protocol as before, using a drive that couples strongly to the BEC's quadrupole mode~\cite{Supplement}.
The relative temperature $T/T_\mathrm{c}$ is varied at a fixed scattering length $a_{\rm BF}\,{=}-400\,a_0$ by reducing the number of bosons while fixing the same final temperature.
The radial trap ellipticity~\cite{Supplement} allows us to distinguish the collisionless mode of the bosons' thermal component at $2\omega_y^{\text{B}}$ from the superfluid breathing mode at $2\omega_{\perp}^{\text{B}}$, as depicted in Fig.~\ref{fig:temp}a.
Fig.~\ref{fig:temp}b
displays the change of the bosons' and fermions' spectral responses with temperature.
At low temperatures, the bosons exhibit resonances at the BEC quadrupole ($\sqrt{2}\omega^\mathrm{B}_\perp$) and breathing ($2\omega^\mathrm{B}_\perp$) modes.
At $T\,{>}\,0.5\,T_\mathrm{c}$ we observe an additional 
peak at $2\omega_y^{\text{B}}$, corresponding to the intraspecies collisionless mode of the bosonic thermal component~\cite{Mewes1996,Jin1996}.
With increasing $T/T_\mathrm{c}$ the bosonic response at the BEC superfluid modes is reduced while the response at the intraspecies collisionless mode grows, until above $T_\mathrm{c}$ only the collisionless mode persists.

The fermions can be viewed as a highly sensitive probe for the complex crossover of modes in the Bose gas.  
At the coldest temperatures, they respond exclusively at the frequencies of the BEC's hydrodynamic modes.
%
Remarkably, starting at $T/T_\mathrm{c}\,{\approx}\,0.4$, two additional modes appear in the fermionic response.
The lower resonance at $2\omega^\mathrm{B}_y$ coincides with the intraspecies collisionless mode of the bosonic thermal component whereas the higher one at $2\omega^\mathrm{F}_y$ coincides with the native collisionless mode of the fermions.
Above $T_\mathrm{c}$, the fermions respond only at their collisionless mode at $2\omega^\mathrm{F}_y$, and this regime is well captured by solving the full coupled Boltzmann equations for the mixture with nonvanishing $I_{\rm coll}$~\cite{Supplement,theory_paper}.
We note that while the thermal components of the bosons and the spin-polarized fermions are both in the collisionless regime by themselves, strong interspecies interactions bring this mixture into local equilibrium~\cite{Ferlaino2003}.
We interpret the fermionic response at the bosonic collisionless mode ($0.3\, T_{\rm c}\lesssim T \lesssim T_{\rm c}$) as an indication of the mixture's crossover from the collisionless to the collisionally hydrodynamic regime.
Above $T_\mathrm{c}$, the mixture reverts to collisionless flow for both species due to the lowered density of bosons.

\begin{figure}
	\includegraphics[width=\columnwidth]{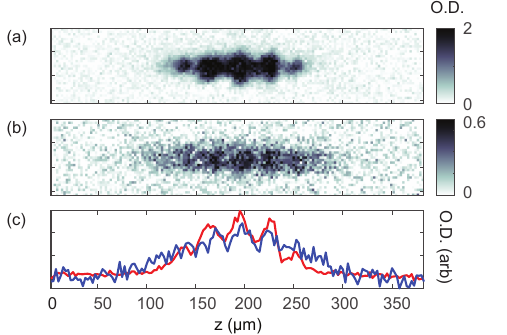}

	\caption{\label{fig:faraday} Faraday waves in a Bose-Fermi mixture (\textit{in-situ}), showing the longitudinal density pattern in the BEC (a) and fermions (b), at an interspecies interaction of $500\,a_0$.
	Bose-Fermi interactions cause the BEC density striation to be imprinted on the fermions. Line densities for the bosons (red) and fermions (blue) are shown in (c), with the fermions rescaled by a factor of three for easier comparison.
	}
	
\end{figure}

The quadrupole and breathing modes are low-lying collective excitations of the coupled system, but perhaps the most visually striking observation of synchronized flow is the appearance of high-order excitations -- Faraday waves~\cite{Faraday1831} -- in the Bose-Fermi mixture.
They arise from the parametric excitation of collective modes transverse to the direction of drive.
In the context of atomic gases, Faraday waves have been observed in elongated BEC's~\cite{Engels2007, Groot2015, Nguyen2019} upon modulating weakly-interacting BECs along the radial direction, inducing striated density patterns along the longitudinal direction.
Strikingly, when inducing Faraday waves in the BEC, we here observe the emergence of Faraday waves also on in the gas of fermions at $a_{\rm BF} = 500\,a_0$(see Fig.~\ref{fig:faraday}). 
For this, we modulated the radial optical potential for 8 cycles at $2\omega^B_r$.
From the observed period of $\lambda_{\rm Far} = 28(7)\,\mu$m in the density striation and the drive frequency we infer the condensate's speed of sound $c=5.5(1.4)\,\mu$m/ms, which is consistent with the Bogoliubov speed of sound obtained from the measured chemical potential $\mu$, $c=\sqrt{\mu/m_B}=4.9(2)\,\mu$m/ms.
To our knowledge, this is the first observation of spatial patterns analogous to Faraday modes observed in a gas of fermions.


Our results demonstrate a novel regime of collective motion of fermions, tracing the superfluid hydrodynamic flow of a Bose condensate.
As the temperature is increased, incoherent collisions between the thermal bosons and fermions cause a crossover into the collision-dominated hydrodynamic regime, in direct analogy to 2D electron gases, where the electron mean-free path is tuned with the density and temperature.
At temperatures lower than those achieved in this study, induced fermion-fermion interactions~\cite{Heiselberg2000,Bijlsma2000} are predicted to arise within the Bose-Fermi mixture, a precursor of the long-sought $p$-wave superfluidity of fermions mediated by bosons~\cite{Efremov2002,Matera2003,Kinnunen2018}.

We acknowledge Eric Wolf for helpful discussions.  We acknowledge support from NSF, AFOSR through a MURI on Ultracold Molecules, the Vannevar Bush Faculty Fellowship. Z.~Z.~Y. and A.~C. acknowledge support from the NSF GRFP.  C.~R. acknowledges support from the Deutsche Forschungsgemeinschaft (DFG) Germany research fellowship (421987027). K.~S. acknowledges funding from NSF EAGER-QAC-QCH award No. 2037687.
P.~D. and E.~D. were supported by ARO grant number W911NF-20-1-0163, the SNSF project 200021-212899. E.~D. acknowledges support from the Swiss National Science Foundation under Division II.
\paragraph{Author contributions}
M.Z. and E.D. conceived of the experiments and supervised the study. Z.Z.Y, Y.N., A.C., and C.R. performed the experiments and the data analysis. Z.Z.Y, C.R., and M.Z. performed the numerical calculations for the Boltzmann equations without collisions and the mean field scaling ansatz.  P.E.D., K.S., and E.D. performed the theoretical and numerical calculations on the high temperature Boltzmann equations.  All the authors contributed to the manuscript and the interpretation of data.
\paragraph{Competing interests: } 
The authors declare no competing interests.

\clearpage

\clearpage
\bibliography{HydroRefs}
\clearpage 
\section{Methods} 
Here we provide further details of the collective oscillation measurements.
The optical potential is generated by two 1064\,nm beams intersecting at 90$^\circ$, oriented in the $z$- and $y$- directions respectively. The trap is   approximately cylindrically symmetric with trap frequencies of $\omega^\mathrm{B}_{x,y,z}/2\pi{=}[103(3), 94(2), 12.2(0.3)]\,$Hz and $\omega^\mathrm{F}_{x,y,z}/2\pi{=}[125(2),114(2),15(1)]$\,Hz for bosons and fermions, which are prepared in their respective hyperfine ground states ($\ket{F = 1, m_F = 1}$ for $^{23}$Na and $\ket{F = 9/2, m_F = -9/2}$ for $^{40}$K).
The fermions are moderately degenerate with $T/T_\mathrm{F}$ ranging from 0.6 to 2, where $T_\mathrm{F}$ is the Fermi temperature.
The large variation stems from mean-field attraction and repulsion as well as different degrees of three-body inelastic loss that removes the minority fermions when $|a_{\rm BF}|$ is large.
The BEC is weakly interacting with a Bose-Bose scattering length of $a_{\rm BB}\,{=}\,52\,a_0$, where $a_\text{0}$ is the Bohr radius. 
Our lowest temperatures are $T/T_\mathrm{c}\,{\lesssim}\,0.2$ with respect to the BEC critical temperature $T_\mathrm{c}$.

To create an interacting Bose-Fermi mixture, we ramp the magnetic field in between two interspecies Feshbach resonances, allowing continuous tuning of the interspecies $s$-wave scattering length, $a_\mathrm{BF}$. First, we ramp the magnetic field to the zero-crossing of the scattering length at 80.3\,G and wait for the field to stabilize for 5\,ms.  Then, the field is quenched to the final interaction strength within 10\,$\mu$s, and oscillations are initiated.  We take care to feed forward the programming of the magnetic field to compensate for slower drifts of eddy currents on the several ms timescales.  We check the field is stable during the oscillation perturbation sequence by independently measuring it using radiofrequency spectroscopy on the impurity atoms.  
During the oscillations, differential gravitational sag between the species is cancelled using a magnetic field gradient.
To center the two species, we empirically found the correct magnetic gradient by minimizing the lifetime of the K atoms at strong interactions with the Na atoms as a function of gradient value. At the best overlap point, the K atoms underwent the fastest inelastic 3-body
losses with Na atoms.
The density overlap profiles may be found in the supplement section on equilibrium line densities~\cite{Supplement}.

We study the collective modes versus interaction strength at a fixed $T/T_\mathrm{C}\,{\approx}\,0.2$ (see main text Fig.~2) by applying a sinusoidal intensity perturbation on the $z$-axis \SI{1064}{\nano\meter} trap beam with a modulation amplitude of 20\% and a variable drive frequency.  
This method injects energy in the $x$- and $y$- motion of the clouds and primarily excites the transverse breathing mode.  
We study the collective modes versus temperature at a fixed interaction strength (set by the scattering length $a_{\rm BF}\,{=}\,-400\,a_0$) by applying intensity modulations to both the $z$- and $y$-directional \SI{1064}{\nano\meter} beams (see main text Fig.~4). The modulation in each direction is $180^\circ$ out of phase in order to better couple to the transverse quadrupole mode.  The modulation depths are 15\% for both beams. 
In both cases, the clouds are imaged after a duration lasting ten oscillation cycles.
To minimize power broadening of the spectra and maximize the Fourier resolution, the modulation amplitude is minimized and the number of cycles is maximized within the experimental limits of signal-to-noise ratio and lifetime from inelastic collisions.  
Our chosen modulation depth is strong enough that the oscillations slightly deviate from linear response.

To study the evolution of collective modes vs temperature, we reduced the evaporation efficiency of the mixture by ramping the optical potential to its final value more quickly, leading to a similar absolute temperature but reduced boson number.

Fitting of the fermion spectrograms, \textit{i.e.}~in Fig.~1 of the main text, is performed with a phenomenological asymmetric function of the form
\begin{align}
    c(\omega,p_i)=\frac{p_1}{f(\omega,p_2,p_3,p_4)} &e^{-(\omega-p_2)^2/f(p_2,p_3,p_4)^2}\nonumber\\
    \text{with}~f(\omega,p_2,p_3,p_4)&=p_3(1+e^{p_4(\omega-p_2)})^{-1}
\end{align}
with $p_i$ as free fitting parameters.  This function reduces to a Gaussian in the limit $p_4\,{=}\,0$, and only the peak location and width are used in data analysis.  The asymmetry and downshifting of the noninteracting fermion resonance from its expected frequency is due to strong driving of the fermions in an anharmonic trap.

In Fig.~1-4 of the main text, the modulation frequencies are shown as normalized to the boson mean radial trap frequency, $\omega_\perp^{\rm B}$, which is $2\pi\times$\,98 Hz on average.  However, long-term drifts of the trap depth over many experimental repetitions necessitated a different normalization for each interaction strength in Fig.\,2-3. The exact normalization value was determined by fitting a Gaussian to the boson spectrograms at frequencies near $\omega_\perp^{\rm B}$, extracting the frequency of the largest response, and assigning that as the normalization for the boson and fermion spectograms for that particular $a_{\rm BF}$.  This procedure necessarily forces the BEC breathing mode in Fig.\,2 to fall at $2\omega_\perp^{\rm B}$, and falls to the same absolute frequency to within 2\% error for all driving conditions. 

\section{Data availability}
The data that support the plots within this paper and other findings of this study are available as source data files. 
 All other data are available from the corresponding author upon reasonable request.

\end{document}